# High-field Magnetotransport Studies of Surface Conducting Diamonds

Kaijian Xing,[1, 2, *] Daniel L. Creedon,[3] Golrokh Akhgar,[4, 5] Steve A. Yianni,[2] Jeffrey C. McCallum,[3] Lothar Ley[2, 6] Dong-Chen Qi [7, 8, ‡], Christopher I. Pakes,[2, †]

[1] *School of Physics and Astronomy, Monash University, Clayton, VIC 3800, Australia*

[2] *Department of Chemistry and Physics, La Trobe Institute for Molecular Science, La Trobe University, Bundoora, Victoria 3086, Australia*

[3] *School of Physics, The University of Melbourne, Victoria 3010, Australia*

[4] *Department of Materials Science and Engineering, Monash University, Clayton, Victoria 3800, Australia*

[5] *ARC Centre of Excellence in Future Low-Energy Electronics Technologies, Monash University, Melbourne, Victoria 3800, Australia*

[6] *Institute for Condensed Matter Physics, Universität Erlangen, Staudt-Str. 1, 91058 Erlangen, Germany*

[7] *School of Chemistry and Physics, Queensland University of Technology, Brisbane, Queensland 4001, Australia*

[8] *Centre for Materials Science, Queensland University of Technology, Brisbane, Queensland 4001, Australia*

**Abstract**

The observation of strong and tunable spin-orbit interaction (SOI) in surface conducting diamond opens up a new avenue for building diamond-based spintronics. Herein we provide a comprehensive method to analyze the magnetotransport behavior of surface conducting hydrogen-terminated diamond (H-diamond) Hall bar devices and $Al/Al_2O_3/V_2O_5$/H-diamond MOSFETs, respectively. By adopting a significantly improved theoretical magneto transport model, the reduced magnetoconductance can be accurately explained both within and outside the quantum diffusive regime. The model is valid for all doping strategies of surface conducting diamond tested. From this analysis, we find that the orbital magnetoresistance, a classical effect distinct from the SOI, dominates the magnetotransport in surface conducting diamond at high magnetic fields. Furthermore, local hole mobilities as high as 1000 ~ 3000 $cm^2$/Vs have been observed in this work, indicating the possibility of diamond-based electronics with ultra-high hole mobilities at cryogenic temperatures.

*: kaijian.xing@monash.edu
‡: dongchen.qi@qut.edu.au
†: c.pakes@latrobe.edu.au

## I. INTRODUCTION

Diamond, a wide bandgap semiconductor, is, due to its extraordinary properties, highly regarded as a promising material for future electronic device applications in fields requiring high-power and high-frequency operation. This promise has been fueled by the discovery of an intriguing *p*-type surface conductivity on diamond when its surface is both hydrogen-terminated and exposed to atmospheric or high-electron-affinity solid-state surface acceptors [1-6]. As a result, the conductive diamond surface has become a fertile arena for developing diamond-based electronics without the need to introduce impurity dopants into the diamond lattice, which suffers from a number of significant drawbacks [7, 8]. Recent improvements in the quality of hydrogen-terminated diamond (H-diamond) surfaces [9] have permitted the surface to support a metallic conductivity even at sub-Kelvin temperatures, from which the two-dimensional (2D) nature of the sub-surface hole accumulation layer has been revealed and the study of quantum transport effects made possible. [10] A number of 2D quantum transport phenomena, such as the Coulomb blockade effect [11] and Shubnikov-de Haas (SdH) oscillations [12, 13] have been reported on surface conducting H-diamond. Using magnetotransport measurements performed in the limit of low applied magnetic field, we have observed quantum coherent backscattering manifested in the form of weak localization (WL) and weak anti-localization (WAL) in the 2D hole conducting layer on diamond introduced both naturally by air [14] and by high work function transition metal oxides (TMOs) as surface acceptors [15]. These studies reveal a surprisingly strong $k^3$-Rashba spin-orbit interaction (SOI) up to 20 meV in the 2D hole band, as a result of the highly asymmetric quantum confinement potential at the surface [16], which can be further modulated via both ionic liquid [17] and solid-state electrostatic gating [18] over a large dynamic range.

Previous magnetotransport measurements on surface conducting H-diamond have been predominantly limited to the quantum diffusive regime (i.e. $|\vec{B}| < 0.1B_{\mathrm{tr}}$) [19] which can be well described by the phase-coherent 2D localization theory. [14, 17] Takahide *et al.* studied magnetotransport on H-diamond and reported a similar positive magnetoresistance at 2 K which persists up to 7 T. They attributed this anomalous magnetoresistance to the spin properties of the holes accumulated at the diamond surface but without modelling this effect [20]. The observation of a positive magnetoresistance well above the diffusive limit is particularly intriguing as it hints that the high-field spin transport cannot be solely described by the weak antilocalization effect. In this work, the magnetoconductance of air- and transition-metal-oxide (TMO)-doped surface conducting diamond Hall bar devices was studied at high applied magnetic fields and analyzed by considering both the classical orbital magnetoresistance and the quantum mechanical phase-coherent backscattering within the $k^3$ Rashba theory. We show that the orbital magnetoresistance is present in surface conducting diamond and dominates the magnetotransport at high magnetic fields. Our analysis further reveals the possible existence of local low carrier density regions at diamond surfaces, consistent with the previously reported spatial inhomogeneity of the 2D hole gas on surface conducting diamonds. [15, 21, 22] Moreover, the mobilities of local holes have been demonstrated to be as high as several thousand $cm^2/Vs$ at low temperatures (down to 0.25 K) which are one order magnitude higher than those reported for surface conducting diamond at low temperature [13, 14, 17, 23]. By improving the interface quality between diamond and surface acceptors, it is possible to achieve a high-mobility 2D metallic Fermi liquid at the diamond surface, promising a rich field of potential future studies such as the quantum Hall effect or potential emergent superconducting states.

## II. EXPERIMENTS

### *2.1 Device fabrication*

Three different H-diamond device structures were fabricated to investigate the high-field magnetoconductance behavior at cryogenic temperatures: air-doped H-diamond Hall bars, TMOs-doped (using $MoO_3$ and $V_2O_5$) H-diamond Hall bars with and without solid-state gating. Four synthetic single-crystalline (100) diamond samples (*A*, *B*, *C* and *D*) were used in this study. After standard cleaning procedures each sample was exposed to a hydrogen plasma for 10 minutes while mounted on a molybdenum sample stage heated to 850 °C in order to achieve the formation of hydrogen termination. Standard photolithography methods used in our previous work were followed to define the Hall bar channels with Pd ohmic contacts on each of the H-diamonds as shown in Fig. S1. (more details in Supplemental Material). [5] Diamond *A* was left in an atmospheric environment for a few days to maximize surface transfer doping from adsorbed water vapor in the air. In order to engineer the TMO-doped H-diamond interface, diamonds *B* and *C* were annealed in an ultra-high vacuum environment at 400 °C for 30 minutes to drive away the atmospheric surface acceptors, followed by the thermal deposition of 20 nm $MoO_3$ (Sample B) and 20 nm $V_2O_5$ (Sample C) *in-situ* through a shadow mask (Fig. S1) onto the Hall bar channels. To fabricate the gated Hall bar device (Sample D), a 5 nm $V_2O_5$ adlayer was first deposited on the surface of H-diamond, followed by the growth of 10 nm of $Al_2O_3$ by atomic layer deposition (ALD) at 200 °C, acting as the gate dielectric. The low ALD reaction temperature was used to protect the hydrogen termination during the ALD process. Finally, an aluminum gate was formed by standard photolithography and lift-off processes as illustrated in Fig. S1.

*2.2 Magnetotransport measurements*

A Leiden Cryogenics dry dilution refrigerator with an integrated 9-1-1 T superconducting vector magnet was employed to undertake magnetotransport measurements in the temperature range 0.25 K to 20 K. The longitudinal resistivity, $\rho_{xx}$, and the Hall resistance, $R_{xy}$, were measured using standard AC lock-in amplifier techniques and an Agilent 34410A multimeter. A Keithley 2450 Source Measure Unit was used to supply a gate voltage to modulate the hole carrier density in the $Al/Al_2O_3/V_2O_5$ gated H-diamond device (Sample *D*).

## III. RESULTS AND DISCUSSION

*3.1 Orbital Magnetoresistance at high magnetic field*

WL and WAL effects, which arise from phase-coherent backscattering, have been observed previously at cryogenic temperatures in surface-conducting diamond. [14, 15, 17, 18] The transition from WL to WAL is strong evidence of SOI in surface conducting diamond, which is caused by the external electric field across the interface between diamond surface and surface acceptors. The asymmetry of the quantum well at the interface causes a Rashba-type SOI at the surface of H-diamond. [16] Following the measurement methodologies used in our previous works, the reduced magnetoconductance ($\Delta\sigma_{xx} = \sigma_{xx}(B) - \sigma_{xx}(B = 0\text{ T})$) in units of $G_0 = e^2/\pi h$ for Sample $A$ (air-doped device) has been plotted as a function of magnetic field at selected temperatures after removal of a contribution to the conductivity arising from hole-hole interactions [14, 24]. The magnetoconductance at low applied magnetic fields, within the quantum diffusive regime [19], exhibits the characteristic behavior of WL and WAL (Fig.1 (a) to (c)): a positive magnetoconductance (PMC) is observed at 4 K but with a characteristic WAL cusp as $B \rightarrow 0$ T that becomes more pronounced as the temperature is reduced. Intriguingly, the WL is also suppressed at high magnetic fields, where a negative magnetoconductance (NMC) is observed at all temperatures. This behavior is different from that reported for surface conducting diamond by Takahide *et.al.* in 2016 [20] and Sasama *et.al.* in 2019 [13], where NMC was only observed in an applied magnetic field (up to 7 T in Takahide's work and 17 T in Sasama's work) down to 1 K. These authors suggested that the effect arose from a spin-induced effect, or the normal classical orbital effect. The trend of NMC→PMC→NMC with increasing field, as shown in Fig. 1 (a), (c) and (d), is strong evidence of the influence of multiple effects in this 2D hole band: the existence of WL and WAL induced by phase coherent backscattering at low-magnetic field, as well as the regular, Drude-type magnetoresistance at high-magnetic field. As reported by Akhgar *et. al.* in 2016, WAL is observed when a high spin-orbit interaction (SOI) is present and the resulting magnetoconductance can be described using the $k^3$ Rashba model (Eq.1) within the quantum diffusive regime: [25-27]

$$\frac{\Delta\sigma_{xx}}{G_0} = \begin{array}{l} \psi\left(\frac{1}{2}+\frac{B_\phi + B_{SO}}{B}\right) + \frac{1}{2}\cdot\psi\left(\frac{1}{2}+\frac{B_\phi + 2\cdot B_{SO}}{B}\right) - \frac{1}{2}\cdot\psi\left(\frac{1}{2}+\frac{B_\phi}{B}\right) \\ - ln\left(\frac{B_\phi + B_{SO}}{B}\right) - \frac{1}{2}\cdot ln\left(\frac{B_\phi + 2\cdot B_{SO}}{B}\right) + \frac{1}{2}\cdot ln\left(\frac{B_\phi}{B}\right) \end{array} \qquad \text{Eq.1,}$$

where $\psi$ is the digamma function. $B_{so}$ and $B_\phi$ represent the characteristic spin- and phase-coherent fields. This model has been shown to reproduce accurately experimental magnetocondutance data for surface conducting diamond devices at low magnetic fields [14, 17], in the range -1 T to 1 T, as illustrated in Fig.1 (d) by the solid blue curve which is derived from a theoretical fit using Eq.1. However, if the fitted curve based on Eq.1 is extended to higher magnetic fields, it deviates significantly from the measured data as shown in Fig. 1(d) suggesting a different mechanism that dominates the high-field magnetotransport behavior. Several studies of 2D electron systems have suggested that the classical orbital magnetoresistance plays dominant role at high magnetic fields, which has been described using a correction to the Drude conductivity of the form $\sigma(B) = {AB^2}/{1 + CB^2}$, [28-29] where A and C are fitting parameters related to the carrier density and mobility. Naumann *et al.* suggested that the orbital magnetoconductance can be written as $\sigma(B) = {n\cdot e\cdot\mu}/{1+(\mu\cdot B)^2}$, where $e$ is the elementary charge, $n$ and $\mu$ represent the carrier density and mobility, respectively. [31] Recently, quantum oscillations have been explored at the hexagonal boron nitride/H-diamond interface [13] in which a two-carrier model has been utilized in the modeling of the magnetoconductance. While this work has not paid attention to the low-field magnetoresistance behavior, a two-band magnetoconductance model of the form $\sigma_{xx} = {n_1\cdot e\cdot\mu_1}/{[1+(\mu_1\cdot B)^2]} + {n_2\cdot e\cdot\mu_2}/{[1+(\mu_2\cdot B)^2]}$ was used to describe magnetoconductance curves obtained at different gate biases and in applied magnetic fields of up to 16 T. In the present work, we further consider the presence of spatial inhomogeneity of the 2DHG in surface-conducting H-diamond, both in terms of local density and mobility [15, 21, 22]. We utilize a two-carrier model as discussed above which yields the following expression for $\Delta\sigma_{orbital}$

$$\Delta\sigma_{\mathrm{orbital}} = \frac{n_1 \cdot e \cdot \mu_1}{1+(\mu_1 \cdot B)^2} - n_1 \cdot e \cdot \mu_1 + \frac{n_2 \cdot e \cdot \mu_2}{1+(\mu_2 \cdot B)^2} - n_2 \cdot e \cdot \mu_2 \quad \text{Eq.2,}$$

where $n_1$ and $\mu_1$ are global carrier density and mobility, representing spatially averaged transport parameters for the device as a whole. $n_2$ and $\mu_2$ represent the average density and mobility in areas of the surface where it may deviate significantly from the Hall parameters. The contribution to the magnetoconductance of $\Delta\sigma_{orbital}$ is very small within the quantum diffusive regime, relative to the WL and WAL corrections (More details are discussed in Supplemental Material, shown in Fig.S2), so Eq.1 and Eq.2 can be simply combined to model the data over the full range of applied magnetic fields:

$$\frac{\Delta\sigma_{xx}}{G_0} = \begin{matrix} \psi\left(\frac{1}{2} + \frac{B_\phi + B_{SO}}{B}\right) + \frac{1}{2} \cdot \psi\left(\frac{1}{2} + \frac{B_\phi + 2 \cdot B_{SO}}{B}\right) - \frac{1}{2} \cdot \psi\left(\frac{1}{2} + \frac{B_\phi}{B}\right) \\ - \ln\left(\frac{B_\phi + B_{SO}}{B}\right) - \frac{1}{2} \cdot \ln\left(\frac{B_\phi + 2 \cdot B_{SO}}{B}\right) + \frac{1}{2} \cdot \ln\left(\frac{B_\phi}{B}\right) + \Delta\sigma_{\mathrm{orbital}} \end{matrix} \quad \text{Eq.3,}$$

To fit the magnetoconductance data in the full magnetic field range, the experimental data within the quantum diffusive regime (-1 T to 1 T) was first fitted with Eq.1 to extract the values of the parameters $B_{so}$ and $B_\phi$. Secondly, using the values of $B_{so}$ and $B_\phi$ so obtained, we fit the magnetoconductance data to Eq.3 and the Hall resistance data to Eq. S1 simutaneously over the full magnetic field range (-7 T to 7 T) by using $n_1$, $\mu_1$, $n_2$ and $\mu_2$ as new fitting parameters. Fits to Eq.3 using this approach adequately describe the experimental data over the full magnetic field range, as illustrated by the solid red curve in Fig.1 (d). The fitted curves derived from Eq.1 and Eq.3 are consistent within the quantum diffusive regime, as expected because the orbital magnetoresistance has minimal effect on the magnetocondutance curves at low field and does not effect the WAL and WL. Consequently, inclusion of the two-carrier orbital magnetoresistance term in the analysis provides an accurate description of the magnetoconductance behaviour over the full range of magnetic fields. The new fitting parameters, $n_2$ and $\mu_2$, derived from this approach will be discussed in detail in the following sections for a variety of surface transfer doping scenarios.

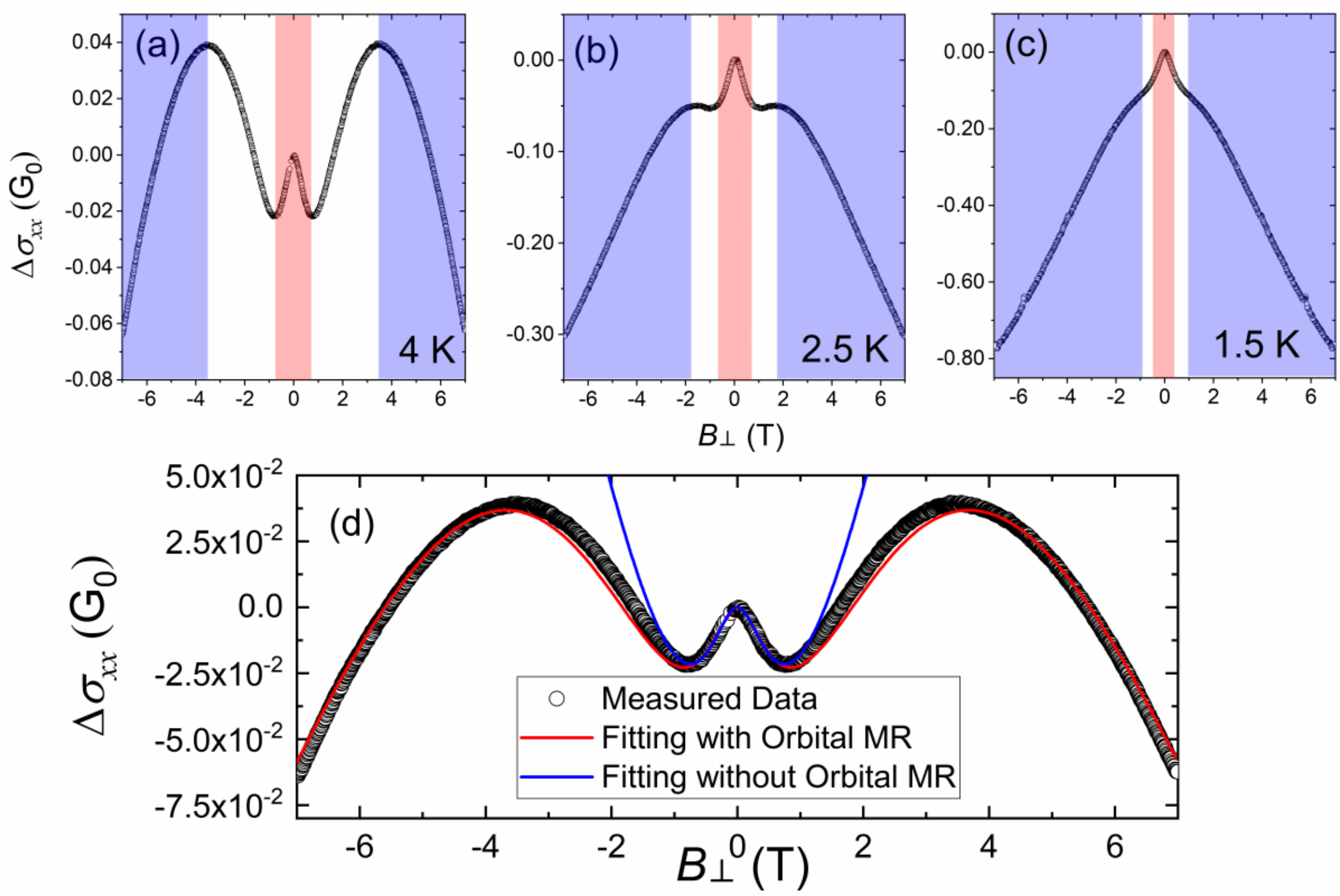

Fig. 1 Reduced longitudinal magnetoconductivity, $\Delta\sigma_{xx}$, in units of $G_0$ at: (a) 4 K; (b) 2.5 K; (c) 1.5 K for the air-doped H-diamond system (sample A). The red regions represent the NMC due to the WAL; the white regions represent the PMC due to the WL; the purple regions represent the NMC due to the orbital magnetoresistance. (d) Fits to theory for reduced longitudinal magnetoconductivity at 4 K the; open circles represent the measured data; the blue line represents the fitting result based on the Rashba model without orbital magnetoresistance; the red line represents the fitting results based on the Rashba model with orbital magnetoresistance. The low magnetic field (-1 T to 1 T) data are taken from *Nano Letters* 2016. [17]

### *3.2 Air-doped H-diamond devices*

For the air-doped H-diamond Hall bar (Sample *A*), magnetotransport measurements were performed as a function of temperature and in an applied magnetic field up to 6 T. As expected, the characteristic WAL cusp developed as the temperature was reduced. By fitting the data within the quantum diffusive regime with Eq.1, $B_{so}$ and $B_{\phi}$ were extracted and are plotted as a function of temperature in Fig.2 (c). The value of $B_{so}$ obtained is 0.29 ± 0.02 T and is almost independent of the sample temperature, while $B_{\phi}$ is observed to increase linearly with temperature, consistent with Nyquist dephasing. [17] As discussed above, Eq.1 becomes invalid for magnetic fields exceeding 1 T due to the existence of the orbital magnetoresistance effect (Fig.2 (a)). Following the new fitting procedure noted above, the values obtained for $B_{so}$, $B_{\phi}$were used with Eq.3 to fit the magnetocondictance curves over the full range of magnetic field at differenct temperatures. At the same time, $n_1/\mu_1$ and $n_2/\mu_2$ were used with Eq.S1 to fit the Hall resistance over the full range of magentic field at different temerpatures (Fig. S3). These yields the solid curves shown in Fig.2 (b) and Fig. S3, which nicely fit all experimental magnetoconductance and Hall resistance data over the entire temperature and magnetic field range. The values of $n_2$ and $\mu_2$ can then be extracted from the fitting procedure as shown in Fig.2 (d) which are significantly different to the extracted $n_1$ and $\mu_1$ values (see Fig. S3 in Supplemental Material). We will return to discuss the values of $n_2$ and $\mu_2$ later.

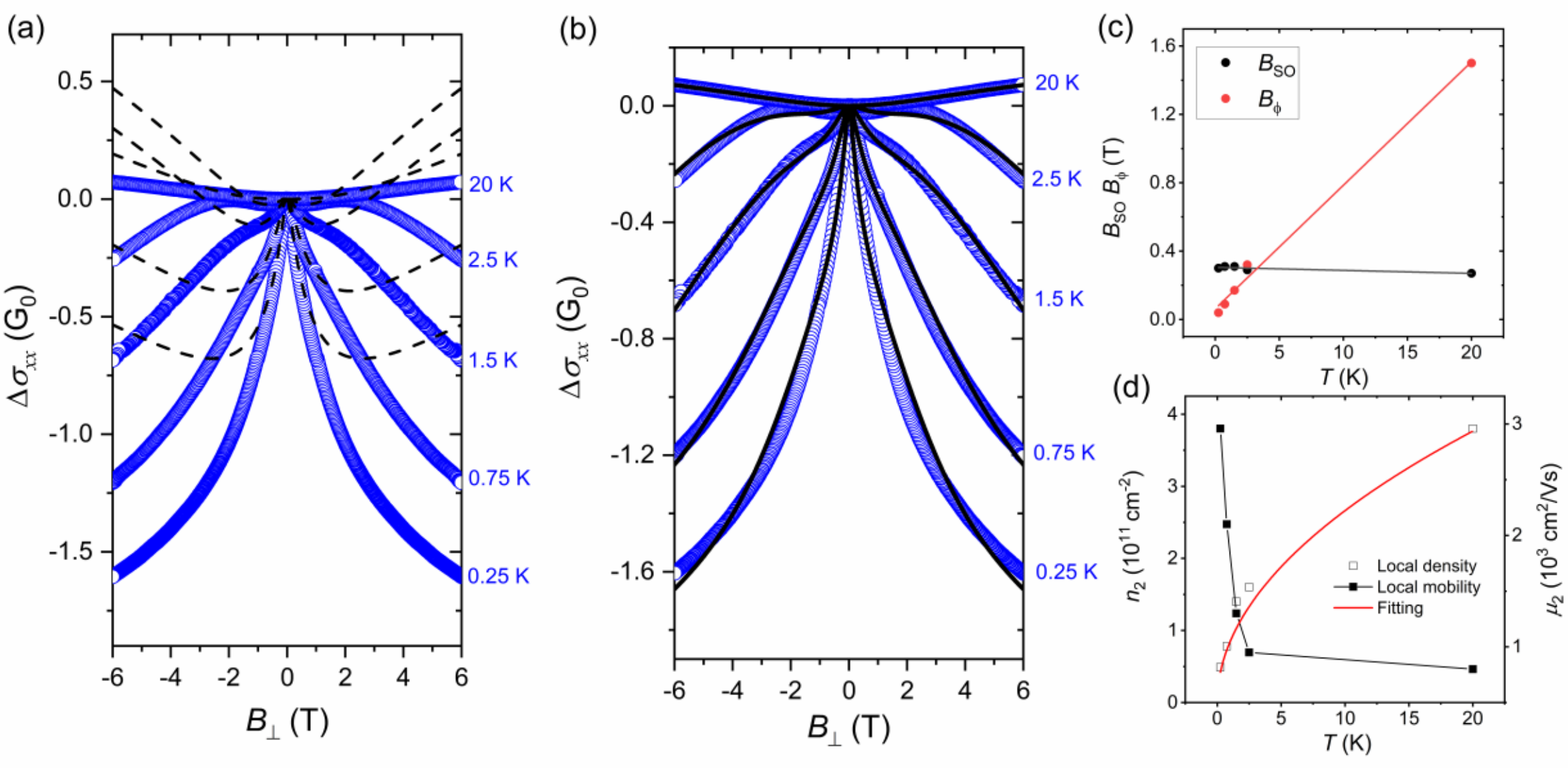

Fig.2 Fits of the reduced magnetoconductance at different temperatures for the air-doped H-diamond system: the open circles represent the measured experimental data; the dashed lines in (a) represent the fits based on the $k^3$ Rashba model without orbital magnetoresistance, whereas the solid lines in (b) represent the fits based on the $k^3$ Rashba model with orbital magnetoresistance. (c) temperature dependence of the characteristic spin-coherence field, $B_{so}$, and phase-coherence field, $B_{\Phi}$; (d) temperature dependence of the local carrier densities (open squares) and carrier mobilities (solid squares) extracted from the new fitting procedure. The low field data (-1 T to 1 T) are taken from *Nano Letters* 2016. [17]

*3.3 TMO-doped H-diamond devices*

We have further investigated the influence of the orbital magnetoresistance on H-diamond devices doped with TMOs. Fig.3 (a) shows the experimental magnetoconductance curves for a $MoO_3$-doped H-diamond device (Sample $B$) fit with Eq.1 within the diffusive region ($|\vec{B}| <$ 0.5 T), with the corresponding fitting parameters $B_{so}$ and $B_{\phi}$ as a function of temperture shown in Fig. 3 (c); the behavior is consistent with our previously reported work. [15] In this case, similarly to the air-doped device (Sample $A$), the fits to theory obtained using Eq.1 deviate significantly from the experimental data for magnetic fields exceeding 1 T. Using the values of $B_{so}$ and $B_{\phi}$ extracted from the low-field fitting process fits to the experimental data using Eq.3 and Eq. S1 are found to satisfactorily describe the experimental magnetoconductance (Fig. 3(b)) and Hall resistance (Fig. S4) curves over the full field and temperature range. Similar analysis was also performed on the $V_2O_5$ doped device (Device $C$) (Fig. S5 and Fig. S8), yielding very similiar results to those shown in Fig.3 and Fig. S4 for the $MoO_3$-doped H-diamond device.

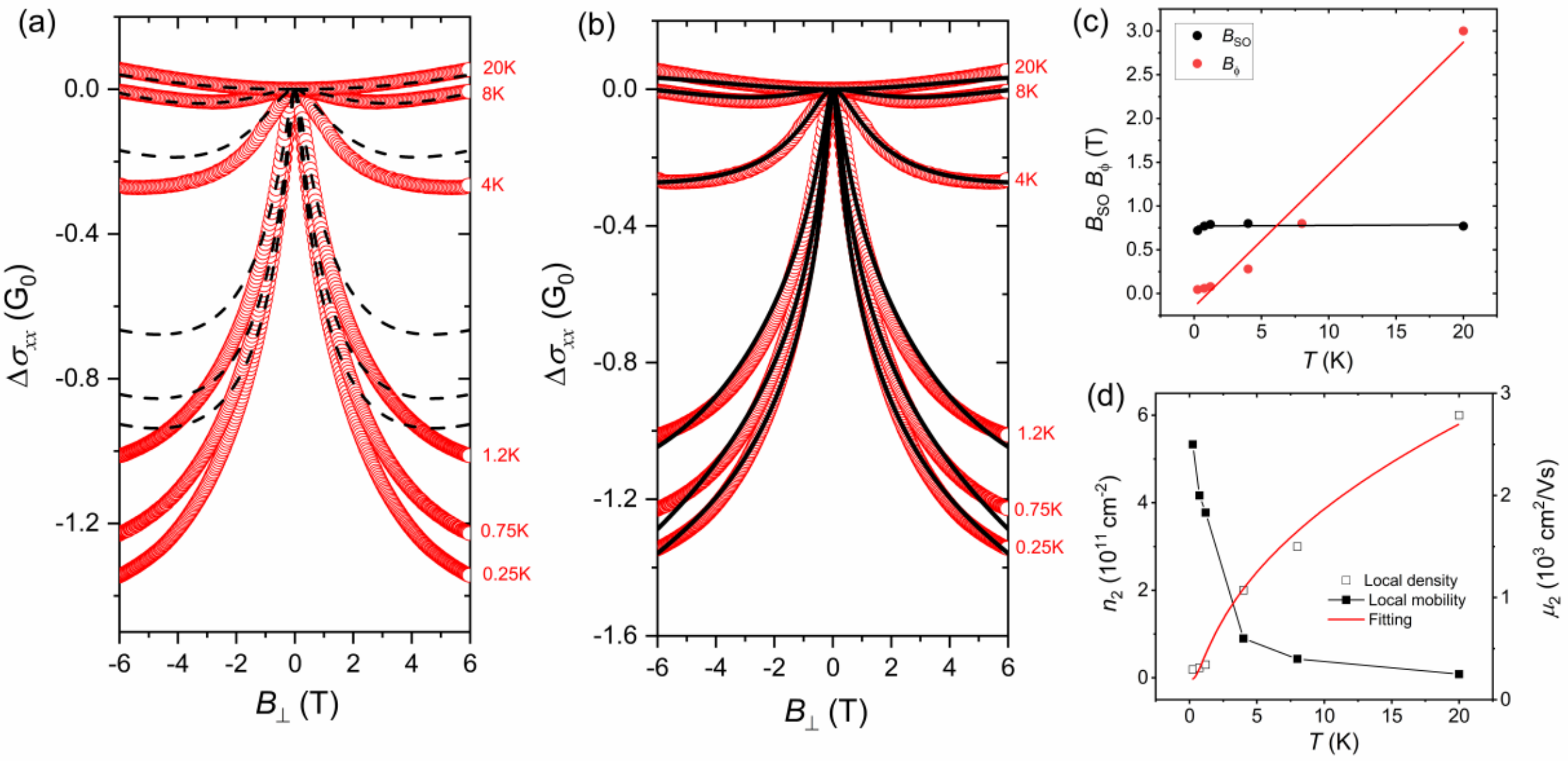

Fig.3 Fits of the reduced magnetoconductance at different temperatures for the $MoO_3$-doped H-diamond system: the open circles represent the measured data; the dashed lines in (a) represent the fits based on the $k^3$ Rashba model without orbital magnetoresistance, whereas the solid lines in (b) represent the fits based on the $k^3$ Rashba model with orbital magnetoresistance. Parameters obtained from the fits for the $MoO_3$-doped H-diamond system: (c) temperature dependence of the characteristic spin-coherence field, $B_{so}$, and phase-coherence field, $B_{\Phi}$; (d) temperature dependence of the local densities (open squares) and mobilities (solid squares) extracted from the new fitting procedure. The low magnetic field data (-1 T to 1 T) are taken from *Carbon* 2020. [15]

*3.4 Local Densities and Mobilities*

The hole density ($n_2$) and mobility ($\mu_2$) derived from fits to Eq.3 and Eq.S1, which we refer to above as the local transport parameters, are plotted as a function of temperature in Fig.2 (d) for Sample *A* and Fig.3 (d) for Sample *B*. At 20 K, the average local hole density $n_2$ is $3.8 \times 10^{11}$ $cm^{-2}$ for the air-doped H-diamond device, and $6.2 \times 10^{11}$ $cm^{-2}$ for the $MoO_3$-doped device, which is approximately two orders of magnitude lower than the global hole densities derived from the Eq.S1 (Fig. S7). This observed reduction in $n_2$ for both cases suggests that localised areas exist on the samples in which the hole density is significantly lower than the macroscopic global Hall density. We argue that this difference is a consequence of the spatial inhomogeneity of the 2DHG at the H-diamond surface which, as noted above, has been suggested from a number of electronic measurements on H-diamond surfaces at low temperatures. [12, 21] Surface roughness, inhomogeneity in the hydrogen termination and the concentration of surface acceptors for example due to the presence of residual photoresist can all give rise to such a spatial inhomogeneity. The local hole density, $n_2$, for air and $MoO_3$-doped devices exhibit a similar dependence on temperature are shown in Fig.2 (d) and Fig.3 (d) respectively; $n_2$ decreases to $4.9 \times 10^{10}$ $cm^{-2}$ at 0.25 K for air-doped H-diamond and to $1.9 \times 10^{10}$ $cm^{-2}$ at 0.25 K for the $MoO_3$-doped device. In Fig.2 (d) and Fig.3 (d) the temperature dependence of $n_2$ was fitted to a function of the form

$$n_2(T) = \sqrt{2k_0 T\varepsilon_r\varepsilon_0 N_v/e^2} \cdot \sqrt{1 + u_s/k_0T + (8/15\pi) \cdot (u_s/k_0T)^{5/2}} \quad \text{Eq. 4}$$

that applies to a degenerate semiconductor. [32] Here $\varepsilon_0$ is the vacuum permeability, $\varepsilon_r$ is the dielectric constant of the diamond ($\varepsilon_{diamond} = 5.7$) and the $MoO_3$ ($\varepsilon_{MoO3} = 9$) [33], $N_v$ is the effective density of states of the diamond valence band and $u_s$ represents the difference between the Fermi energy and the diamond valence band edge. It can be written as $u_s = E_f$ – VBM, where $E_f$ represents the fermi level and VBM represents the valence band maximum. [34] As illustrated in the figures, the data for both devices fit Eq. 4 well and they yield a very small positive $u_s$ (3.0 meV for the air-doped device and 1.0 meV for the $MoO_3$ doped devices). The positive value of $u_s$ indicates that the local regions are weakly degenerate while the macro regions reach the strong degenerate doping condition. [35] The observed reduction in $n_2$, as the temperature is reduced is accompanied by an increase in the corresponding mobility, $\mu_2$. At a temperature of 250 mK, the mobility reaches as value of 2950 $cm^2$/Vs for the air-doped device and of 2500 $cm^2$/Vs for the $MoO_3$ doped device. The extracted local mobilities almost reach those of bulk diamond. Similar data were reported by Takahide *et al.* in 2014, [12]. This and the present work demonstrate local high mobility behavior of the 2DHG at the diamond surface, albeit from different perspectives. The underlying reason for the locally high mobility may be due to weaker impurity scattering effects at these local areas. For example, some microscopic areas having fewer surface acceptor ions have, as a consequence, fewer negatively charged scattering centers which enhances the hole mobility.

### *3.5 Al/$Al_2O_3$/$V_2O_5$/H-diamond gated device*

Having explored the dependence of the transport parameters on temperature, we now turn to examine how the local density and mobility can be modulated via electrostatic gating in the H-diamond gated device (Sample *D*). Experimental magnetoconductance and Hall resistance curves obtained at different gate bias, along with fits to theory as described by Eq. 3 and Eq.S1 obtained with different gate biases at 1.2 K, are shown in Fig.4. The local density, $n_2$, derived from fits to theory can be tuned by applying a negative gate bias. It linearly increased from 3.0 $\times 10^{10}$ cm$^{-2}$ at - 1 V, to 1.5 $\times 10^{11}$ cm$^{-2}$ at - 6V, showing a lower tuning dynamic range compared with the macro regions as a function of gate bias; compare Fig. 4 (b). In Fig. 4 (c), with increasing hole density, the hole mobilities linearly drop from 2200 cm$^2$/Vs to 1300 cm$^2$/Vs. However, examining the carrier density as a function of gate voltage, the capacitance of localised areas *versus* the hole channel shows a difference of orders of magnitude as noted above. The capacitance calculated from the Hall density variation is 0.5 μF/cm$^2$, whereas the one of the local areas is only 2.7 nF/cm$^2$, which is hard to explain using a simple capacitor model. This discrepancy at least suggests that the applied voltage cannot uniformly gate the hole layer confined by the potential wells, but the reasons are not immediately clear. A systematic study of these local islands at the diamond surface by specific technologies, such as scanning tunneling microscopy technology, would be valuable future work. Such a local high mobility suggests that further improving the interface between diamond and surface acceptors, such as physically removing the scattering centers [36], will modify the hole mobility performance, which is a key requirement for unlocking the capability for useful high-frequency electronics.

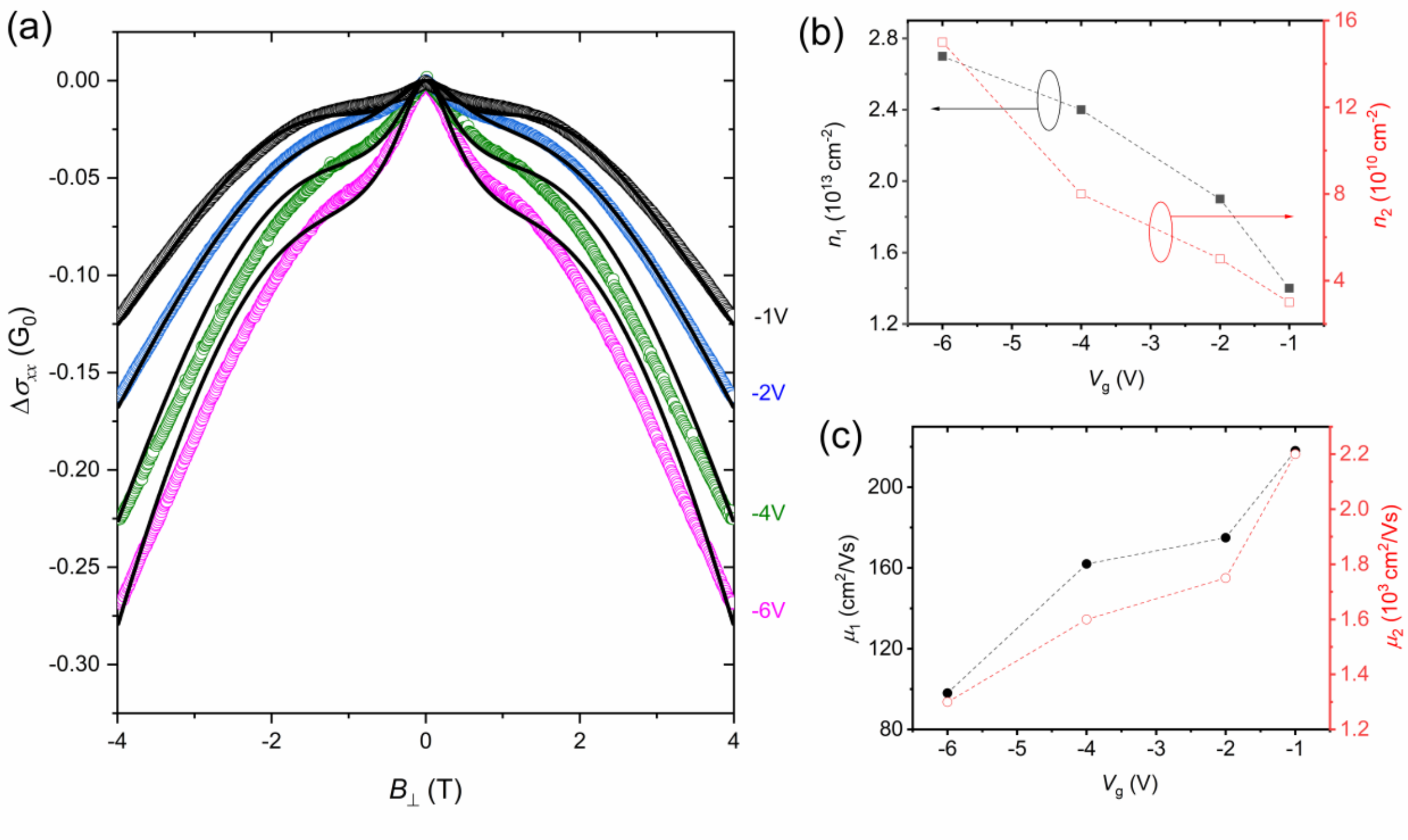

Fig 4. (a) reduced longitudinal magnetoconductance in unit of $G_0$, measured for different gate bias voltages at 1.2 K. Open circles represent the experimental data; the solid lines are fits to Eq. 3. Local carrier densities represented as black squares (b) and local mobilities represented as black circles (c) extracted from the new fitting procedure at different gate bias are shown. The low magnetic field (-1 T from 1 T), Hall densities and mobilities are taken from *Applied Physics Letters* 2020. [18]

## IV. CONCLUSION

In summary, we have presented a comprehensive study of the magnetotransport properties of surface conducting diamond by investigating air-doped and $MoO_3$ (or $V_2O_5$)-doped diamonds, and $Al/Al_2O_3/V_2O_5$/H-diamond MOSFETs. In addition to the previously reported WAL and WL effects on H-diamond surfaces, classical orbital magnetoresistance has been observed and demonstrated to dominate the high-field magnetoresistance at low temperatures. By modifying the theoretical fitting model with an orbital contribution, we provide an exhaustive method to fully understand the magnetotransport behavior of surface conducting diamonds from -6 T to 6 T. In addition, the fitting results, local densities and local mobilities, point towards the existence of local surface regions with low densities in both air-induced and TMO-induced 2DHG systems at diamond surfaces, which is consistent with previously reported spatial inhomogeneity. [15, 21, 22] Moreover, the fitting results also suggest the hole mobilities in these low-density regions are as high as several thousand $cm^2/Vs$ at low temperatures. High mobility is always desirable in surface conducting diamond systems, however the highest reported Hall mobility up to dates is of order several hundred $cm^2/Vs$, which is greatly limited by different scattering mechanisms. [13, 37, 38] By improving the interface quality between diamond and surface acceptors, it appears to minimize the scattering effects and be feasable to achieve a high-mobility 2D metallic Fermi liquid at the diamond surface, which may enable novel experiments investigating emergent superconductivity in this system, or exotic physics such as the quantum Hall effect.

## ACKNOWLEDGEMENTS

This work was supported by the Australian Research Council under the Discovery Project (No. DP150101673). Part of this work was performed at the Melbourne Centre for Nanofabrication (MCN) in the Victorian Node of the Australian National Fabrication Facility (ANFF). D.-C. Q. acknowledges the support of the Australian Research Council (Grant No. FT160100207). D.-C. Q. acknowledge continued support from the Queensland University of Technology (QUT) through the Centre for Materials Science. D.L.C. is supported by Australian Research Council grant DP190102852. K. X. is supported by Australian Research Council grant DP200101345.